\documentclass[prb,aps,superscriptaddress,twocolumn,amssymb,showpacs]{revtex4}

\usepackage{color}
\usepackage{graphicx}
\usepackage{longtable}
\usepackage{epsfig}
\usepackage{dcolumn}
\usepackage{bm}
\usepackage{amssymb}

\begin{document}

\title{      Dimensional crossover and the magnetic transition in electron
             doped manganites }

\author {    Andrzej M. Ole\'s }
\affiliation{Max-Planck-Institut f\"ur Festk\"orperforschung,
             Heisenbergstrasse 1, D-70569 Stuttgart, Germany  }
\affiliation{Marian Smoluchowski Institute of Physics, Jagellonian
             University, Reymonta 4, PL-30059 Krak\'ow, Poland }

\author{     Giniyat Khaliullin }
\affiliation{Max-Planck-Institut f\"ur Festk\"orperforschung,
             Heisenbergstrasse 1, D-70569 Stuttgart, Germany  }

\date{\today}

\begin{abstract}
We introduce a microscopic model for electron doped manganites that 
explains the mechanism of the observed transition from $G$-type 
antiferromagnetic ($G$-AF) to $C$-type antiferromagnetic ($C$-AF) 
order under increasing doping by double exchange mechanism. The model 
unravels the crucial role played by $e_g$ orbital degrees of freedom 
and explains the observed metal-to-insulator transition by a 
dimensional crossover at the magnetic phase transition. 
The specific heat and the spin canting angle found for the $G$-AF 
phase agree with the experimental findings.
As a surprising outcome of the theory we find that spin canting is 
suppressed in the $C$-AF phase, in agreement with the experiment,
due to the Fermi surface topology.

{\it Published in: Phys. Rev. B {\bf 84}, 214414 (2011).}
\end{abstract}


\maketitle

\section{Introduction}
\label{sec:int}

Doped transition metal compounds exibit a number of very interesting 
and fascinating phenomena, including high temperature superconductivity 
in cuprates\cite{Ima98} or pnictides,\cite{Pag10} and colossal 
magnetoresistance in manganites.\cite{Dag01}
The hole-doped perovskite manganites are well
known for their complexity, originating from the competing tendencies 
in their rich phase diagrams,\cite{Tok06} with a variety of spin, 
charge and orbital orders. Several phases are energetically close to 
each other and small changes of electronic parameters or electron
concentration result in phase transitions.\cite{Dag01,Feh04} 
Hole doping, realized e.g. in La$_{1-y}$(Ca,Sr)$_y$MnO$_3$ for 
increasing $y$, triggers then a transition from an antiferromagnetic 
(AF) to a ferromagnetic (FM) phase by the double-exchange (DE) 
mechanism\cite{Zen51} that became a crucial idea to explain the onset 
of the metallic FM phase and electronic transport. 

The DE model is one of the most widespread models of ferromagnetism.
It describes the kinetic energy of electronic charge carriers coupled 
by Hund's exchange to localized $t_{2g}$ spins interacting by 
AF superexchange. A pioneering work by de Gennes\cite{deG60} suggested 
that the competition between FM DE interaction that arises in doped 
systems and AF superexchange between $t_{2g}$ spins leads to 
spin-canted AF order. 
Extensive studies of these ideas led to progress in the theoretical 
understanding of the phase diagrams of hole doped manganites,
\cite{Dag01,Feh04,Mae98,Hot00} while electron doped systems were not 
analyzed carefully until now.
In recent years the DE model was used to study, {\it inter alia\/}, 
striped multiferroic phases in doped manganites,\cite{Don09,Lia11} 
evolution of spin, orbital and charge order at interfaces,\cite{Jac08}
valence fluctuations in magnetite,\cite{McQ07}, novel spin phases
on frustrated lattices,\cite{Mot10} and unuasual spin superstructure 
observed in several vanadium spinels with degenerate $t_{2g}$ 
orbitals.\cite{Che11} In all these cases $e_g$ orbital 
degeneracy plays a crucial role and might lead to anisotropic AF 
phases: either $C$-type AF ($C$-AF) or $A$-type AF ($A$-AF) 
phase,\cite{notecg} depending on the electronic filling and the 
strength of AF superexchange.\cite{vdB99} 

Usually, however, accurate studies of the DE model were hindered by the 
presence of strong intraorbital Coulomb interaction $U$ in a lattice 
distorted by strong cooperative Jahn-Teller effect.\cite{Feh04,Leo10}
A qualitatively different situation arises in an ideal cubic perovskite 
SrMnO$_3$, with $e_g$ bands being completely empty. Here the only 
interaction is AF superexchange between $S=3/2$ spins at Mn$^{4+}$ 
ions. They are generated by $t_{2g}$ charge excitations\cite{Fei99} 
and stabilize the isotropic three-dimensional (3D) 
$G$-type AF ($G$-AF) phase.\cite{notecg} Weakly doped SrMnO$_3$ 
represents an ideal situation for the DE model as the $e_g$ electrons 
hardly interact with each other at low doping $x$ and strong on-site 
Coulomb interactions between them may be neglected. In this case the DE 
model can be almost exactly solved under a single approximation, i.e., 
neglecting weak quantum fluctuations of large $S=3/2$ spins.

Early studies of electron doped Ca$_{1-x/2}$Ce$_{x/2}$MnO$_3$ systems 
gave a magnetic transition to the $C$-AF phase.\cite{Zen01}
The magnetic order in doped CaMnO$_3$ was analyzed within the 
density-functional theory calculations that give instead a transition 
from the $G$-AF to $A$-AF phase.\cite{Tsu10} Recently single crystals 
of Sr$_{1-x}$La$_x$MnO$_3$ and Sr$_{1-x/2}$Ce$_{x/2}$MnO$_3$ were 
synthesized and investigated,\cite{Sak10} showing that the transition 
to the $C$-AF phase occurs in both systems, in agreement with the 
prediction by Maezono {\it et al.\/},\cite{Mae98} and the magnetic 
phase diagram is remarkably universal as a function of doping $x$. 

In electron doped manganites chemical doping of Sr$^{2+}$ with 
La$^{3+}$ (Ce$^{4+}$) ions generates $e_g$ electronic carriers at 
Mn$^{3+}$ ions. Following the standard picture of de Gennes,\cite{deG60}
one expects spin canting on AF bonds. Until recently only one exception 
from this rule was reported --- spin canting is absent in the $A$-AF 
phase when electrons occupy $x^2-y^2$ orbitals and the hopping along AF 
bonds between FM $ab$ planes is blocked.\cite{Mae98} In this context the
recent experimental result of no canting in the $C$-AF phase\cite{Sak10}
is puzzling and requires theoretical explanation, as hopping on the AF 
bonds in $ab$ planes is there finite for electrons within $3z^2-r^2$ 
orbitals, and spin canting is expected.

Hole doped La$_{1-y}$Sr$_y$MnO$_3$ manganites with large $e_g$ electron 
density remain insulating up to $y\simeq 0.17$,\cite{Tok06} as the 
metallic behavior is hindered here by formation of orbital 
polarons.\cite{Kil99,Dag04}. This is in sharp contrast with the 
electron doped systems with the $G$-AF {\it metallic\/} phase found in 
Sr$_{1-x}$La$_x$MnO$_3$ by Sakai {\it et al.\/}\cite{Sak10} already at 
a very low electron doping $x=0.01$. Surprisingly, when doping 
increases further to $x\simeq 0.04$ this 3D weakly metallic phase 
changes to an insulating phase with quasi one-dimensional (1D) 
orbital order of partly occupied $3z^2-r^2$ orbitals, accompanied by 
tetragonal lattice distortion and the $C$-AF spin order. 

In this paper we investigate the DE model for doped $e_g$ electrons, 
extended here by the coupling 
to the tetragonal lattice distortion which occurs in the $C$-AF phase.
We derive the exact electronic structure, show a remarkable difference 
between the Fermi surface topology in the above magnetic phases, and 
demonstrate that it explains the absence of spin canting in the 
$C$-AF phase.
The calculated values of the specific heat coefficient $\gamma$ and 
the critical doping $x_c$ for the $G$-AF/$C$-AF phase transition are
in quantitative agreement with experiment.

The paper is organized as follows. In Sec. \ref{sec:model} we introduce
the microscopic model and specify its parameters. The model is solved 
in Sec. \ref{sec:bands}, where we analyze the remarkable difference
between the electronic structure found in two magnetic phases.  This
leads to the magnetic transition which occurs under increasing doping 
as we show in Sec. \ref{sec:phd}. The paper is summarized in Sec.
\ref{sec:summa}.

\section{The double exchange model}
\label{sec:model}

We consider the DE model for degenerate $e_g$ electrons,\cite{vdB99} 
extended by the coupling to the lattice,
\begin{eqnarray}
{\cal H}&=&\!-\sum_{ij,\alpha\beta,\sigma} t_{\alpha\beta}^{ij} 
c_{i\alpha\sigma}^{\dagger} c_{j\beta\sigma}^{} 
-2J_H\sum_i {\vec S}_i\cdot{\vec s}_i \nonumber \\
&+&\!J\sum_{\langle ij\rangle}{\vec S}_i\cdot{\vec S}_j
- gu\sum_i(n_{iz}-n_{ix})+\frac12 NK u^2.
\label{model}
\end{eqnarray}
Here the $c_{i\alpha\sigma}^{\dagger}$ operator creates an electron with 
spin $\sigma=\uparrow,\downarrow$ in orbital $\alpha=z,x$ at site $i$,
and $n_{i\alpha}\equiv\sum_{\sigma}c_{i\alpha\sigma}^{\dagger}
c_{j\alpha\sigma}^{}$ in electron density in orbital state $\alpha$ at
site $i$, and $N$ is the number of sites. 
Two $e_g$ orbitals are labeled as
\begin{equation}
z\equiv (3z^2-r^2)/\sqrt{6}, \hskip 0.7cm 
x\equiv (x^2-y^2)/\sqrt{2}, 
\label{zx}
\end{equation}
and correspond to two components of the
pseudospin $\tau=1/2$. The hopping elements between a pair of nearest 
neighbors $(ij)$ along the axis $\gamma=a,b,c$ are: 
\begin{equation}
t_{\alpha\beta}^{a/b}=\frac{t}{4}\left( \begin{array}{cc}
1 & \pm\sqrt{3}  \\[0.2cm]
\pm\sqrt{3} & 3
\end{array}\right)\,,\hskip .5cm
t_{\alpha\beta}^{c}=t\left( \begin{array}{cc}
1 & 0  \\[0.2cm]
0 & 0 
\end{array}\right)\,,
\label{tij}
\end{equation}
where $t$ in an effective $(dd\sigma)$ hopping element between two 
$3z^2-r^2$ orbitals along the $c$ cubic axis. Following the 
estimates of $t$ performed within the polaronic picture\cite{Kil99} 
and using the optical spectroscopy for LaMnO$_3$,\cite{Kov10} we fix 
the effective $(dd\sigma)$ hopping element at $t=0.4$ eV.

Hund's exchange between $e_g$ and $t_{2g}$ electrons
at Mn$^{4+}$ ions $J_H\simeq 0.74$ eV is here the largest parameter. 
It aligns the spin of an $e_g$ electron ${\vec s}_i$ with the localized 
$t_{2g}$ spin ${\vec S}_i$ at site $i$. In this way the electronic 
structure depends crucially on the magnetic order of localized $t_{2g}$ 
spins $\{{\vec S}_i\}$ that interact by the nearest neighbor AF 
superexchange $J>0$. It follows from virtual charge excitations
$d_i^3d_j^3\rightleftharpoons d_i^4d_j^2$ along
nearest-neighbor bonds $\langle ij\rangle$ in the
presence of strong on-site 
Coulomb interaction $U$ between $t_{2g}$ electrons.\cite{Fei99}

The tetragonal distortion $u$ is finite only in the $C$-AF 
phase.\cite{Sak10} Here we define it as proportional to a difference 
between two lattice constants $a$ and $c$ along the respective axis, 
$u\equiv 2(c-a)/(c+a)$. Increasing lattice distortion ($u>0$) causes 
increasing $e_g$ orbital splitting $4E_{\rm JT}$ determined by the 
Jahn-Teller energy, $E_{\rm JT}=g^2/2K\simeq 0.2$ eV.\cite{Kil99}
Furthermore, using the experimental data of Ref. \onlinecite{Sak10}, 
we arrived at a semiempirical relation $u\simeq 3x/20$.

\section{Electronic structure}
\label{sec:bands}

Following de Gennes,\cite{deG60} we determined the electronic 
structure for the $G$-AF phase by considering spin canting by angle 
$\theta$ caused by electron doping. Hund's exchange at each site favors
aligned $t_{2g}$ and $e_g$ electron spin states. To solve the 
electronic structure we introduce the fermion 
$\{f_{i\alpha\sigma}^{\dagger}\}$ operators in a canted spin background 
using the local coordinate frame, for instance,
\begin{equation}
c_{i\alpha\uparrow}^{\dagger}=\cos(\theta/2)\,f_{i\alpha\uparrow}^{\dagger}
                             -\sin(\theta/2)\,f_{i\alpha\downarrow}^{\dagger}\,.
\label{fermion}
\end{equation}
The canting of each spin by angle $\theta\in[0,\pi/2]$ (other states 
are equivalent) corresponds for $\theta=0$ $(\pi/2)$ to the AF (FM) 
spin order. The canting modulates the hopping elements between 
neighboring sites, and one finds an analytic solution for two (majority) 
$e_g$ bands in the $G$-AF phase for spins being parallel to the $t_{2g}$ 
spins at each site due to Hund's exchange,
\begin{equation}
\varepsilon_{\bf k}^{\pm}=
-\sqrt{(t\sin\theta\gamma_{\bf k}^{\pm}+J_HS)^2
+(t\cos\theta\gamma_{\bf k}^{\pm})^2}\,,
\label{egaf}
\end{equation}
where 
\begin{eqnarray}
\gamma_{\bf k}^{\pm}&=&\sum_{\lambda}
\cos k_{\lambda} \pm \Big\{\sum_{\lambda}\cos^2 k_{\lambda}
\nonumber \\ 
&-&\cos k_x\cos k_y-\cos k_y\cos k_z-\cos k_z\cos k_x\Big\}^{1/2},
\nonumber \\ 
\label{gamma}
\end{eqnarray}
and the summations in Eq. (\ref{gamma}) are over three cubic 
directions in momentum space, i.e., $\lambda=x,y,z$.
The remaining $e_g$ bands correspond to excited states of $e_g$ 
electrons with spins antiparallel to those of localized $t_{2g}$ 
electrons --- they are given by $-\varepsilon_{\bf k}^{\pm}$ and 
separated from the lower bands (\ref{egaf}) by a large gap of $2J_HS$ 
at $\gamma_{\bf k}^{\pm}=0$. 

\begin{figure}[t!]
\includegraphics[width=8cm]{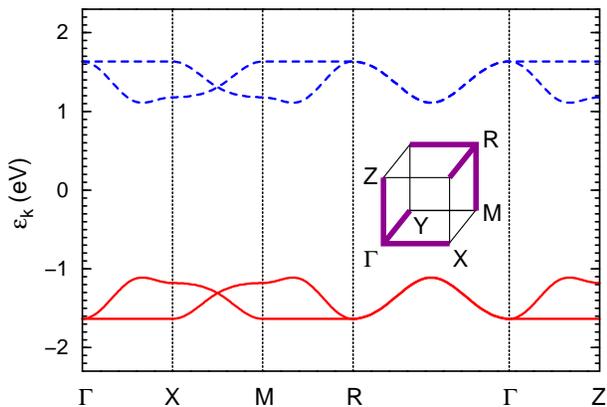}
\caption{(Color online) Band structure in the $G$-AF phase along
the high symmetry directions. Spin majority (minority) bands are
shown by solid (dashed) lines. Parameters: $t=0.4$ eV, $J_H=0.74$ eV, 
$x=0$. Inset shows the Fermi surface at low doping (heavy lines along 
$\Gamma-X$, $R-M$, and equivalent directions) and the special points: 
$\Gamma=(0,0,0)$, $X=(\pi,0,0)$, $M=(\pi,\pi,0)$, $R=(\pi,\pi,\pi)$, 
$Z=(0,0,\pi)$.
}
\label{fig:gaf}
\end{figure}

In the undoped $G$-AF phase of SrMnO$_3$ spin canting is absent 
($\theta=0$) and the electronic structure consists of four Slater 
bands $\pm\varepsilon_{\bf k}^{\pm}$, see Fig. \ref{fig:gaf}. Both 
$e_g$ orbitals are equivalent here and equally contribute to each 
electronic band. The lowest energy is obtained along the $\Gamma-X$ 
and equivalent directions in the Brillouin zone and here electrons are 
doped. By considering the electron energy close to these 
lines one finds that the Fermi surface in the regime of low electron 
doping, shown in the inset in Fig. \ref{fig:gaf}, consists of narrow
cylinders along equivalent main directions in the reciprocal space.

The canting angle in the $G$-AF phase is given by
\begin{equation}
\sin\theta\simeq\frac{tJ_HS}{\sqrt{(3t)^2+(J_HS)^2}}\,\frac{x}{12JS^2}\,,
\label{canting}
\end{equation}
and increases linearly with $x$ in the low doping regime. Using an 
empirical formula for the N\'eel temperature of a Heisenberg 
antiferromagnet,\cite{Fle04} we obtained $J=3.79$ meV for SrMnO$_3$ 
from the experimental $T_N\simeq 230$ K.\cite{Sak10} For the parameters 
of Fig. \ref{fig:gaf}, one finds from Eq. (\ref{canting}) that the 
canting angle is rather small, $\sin\theta\simeq 2.7x$, and quite close 
to the experimental value, $\theta_{\rm exp}\simeq 2x$.\cite{Sak10}

\begin{figure}[t!]
\includegraphics[width=8cm]{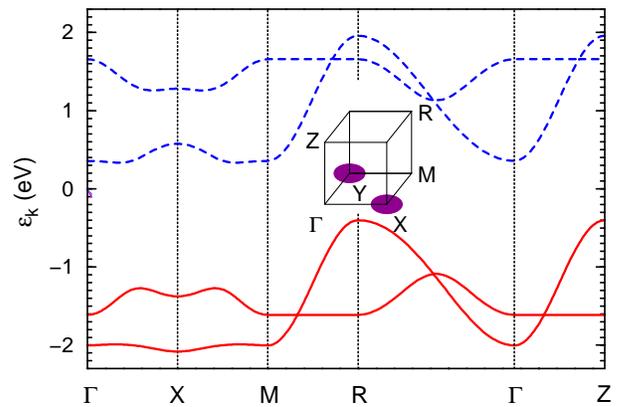}
\caption{(Color online) Band structure in the $C$-AF phase along
the high symmetry directions. Spin majority (minority) bands are
shown by solid (dashed) lines. Parameters: $t=0.4$ eV, $J_H=0.74$
eV, $g=3$ eV, $x=0.05$. The special points as in Fig. \ref{fig:gaf}.
Inset shows the Fermi surface pockets at low doping near the $X$ 
and $Y$ points. 
}
\label{fig:caf}
\end{figure}

The electronic structure $\{\varepsilon_{\bf k}\}$ for the $C$-AF phase 
is quite distinct from the one in the $G$-AF phase and was determined 
numerically. It depends on doping $x$ as the tetragonal lattice 
distortion increases with it. We have verified that 
this dependence is weak for the realistic parameters, i.e., with 
$g\simeq 3.0$ eV and $K\simeq 20$ eV.\cite{Kil99} A representative 
result is shown in Fig. \ref{fig:caf} for $x=0.05$. As the FM order 
along the $c$ axis breaks the cubic symmetry, the orbitals decouple 
from each other in the electronic structure and form two independent
subsets of bands. This is best visible along the $\Gamma-Z$ directions 
where one band has a large dispersion close to $4t$, and the other one 
is dispersionless (Fig. \ref{fig:caf}). In addition, the former band 
has also a very weak dispersion along two (equivalent) $\Gamma-X$ and 
$\Gamma-Y$ directions, with two minima at the $X$ and $Y$ points that 
arise due to the weak mixing between $z$ and $x$ orbitals by hopping in 
$ab$ planes, but this hopping is almost fully quenched due to the AF 
order. 

Surprisingly, we found no spin canting in the $C$-AF phase. Here the
spins form an ideal staggered AF structure in the $ab$ planes, while
they are aligned in FM chains along the $c$ axis. The reason of this 
behavior is the electronic structure with the lowest energy at the $X$ 
and $Y$ points and the Fermi surface consisting of two pockets that are 
centered around these two equivalent points at low doping, see the 
inset in Fig. \ref{fig:caf}. For this Fermi surface topology no kinetic 
energy can be gained by spin canting, since the resulting 
nearest-neighbor hopping between $z$ orbitals in $ab$ planes, 
$\frac14 t\sin\theta(\cos k_x+\cos k_y)$, vanishes at the $X$ and $Y$ 
points. Thus, the observed suppression of spin canting in the $C$-AF 
phase follows from the abrupt change of the Fermi surface topology at 
the magnetic transition.

The consequences of flavor separated band structure in the $C$-AF phase
are remarkable. First of all, electrons doped in SrMnO$_3$ have 
pure $z$ orbital character. Second, this implies that they can 
propagate only along the $c$ axis where localized spins are FM, but 
their motion is blocked in the $ab$ planes by the AF order, 
in agreement with DE. Altogether this leads to a dimensional reduction 
of electronic transport to 1D chains along the $c$ axis.

\begin{figure}[t!]
\includegraphics[width=8.2cm]{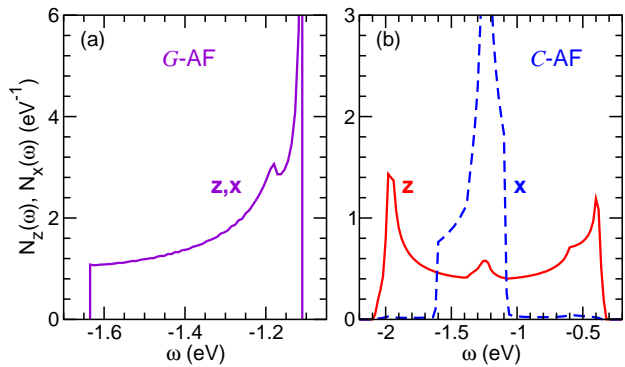}
\caption{(Color online) 
Densities of states for $e_g$ electrons, $N_z(\omega)$ and 
$N_x(\omega)$, in the spin majority bands as obtained for:
(a) the $G$-AF phase ($N_z(\omega)$ and $N_x(\omega)$ are here 
degenerate) with a singularity arising from the AF order at 
$\omega=-1.11$ eV, and
(b) the $C$-AF phase; here the density of $z$, $N_z(\omega)$, and $x$,
$N_x(\omega)$, electron states is shown by solid (red) and dashed 
(blue) line. Parameters: $t=0.4$ eV, $J_H=0.74$ eV; for $C$-AF phase 
in addition $g=3$ eV and $x=0.05$. }
\label{fig:dos}
\end{figure}

A qualitatively different character of the electronic structure in both 
AF phases is visible in the orbital-resolved density of states (DOS) 
$N_z(\omega)$ and $N_x(\omega)$ in the spin majority bands, see Fig. 
\ref{fig:dos}. In the $G$-AF phase both orbital DOSs are the same and 
have low bandwidth of $\sim 0.5$ eV due to the AF order in all three 
cubic directions [Fig. \ref{fig:dos}(a)]. At the lower band edge one 
finds a characteristic step which reflects the quasi-two-dimensional 
DOS of $e_g$ electrons.\cite{Fei05} A large gap between the spin
majority and spin minority states in the AF phase generates a van Hove 
singularity at the upper edge of the DOSs.

From the DOS in the $G$-AF phase, see Fig. \ref{fig:dos}(a), we have 
estimated the coefficient $\gamma\simeq 5$ mJ/mol$\cdot$K$^2$ in the 
specific heat $c_V\simeq\gamma T$ at $x=0.01$ which is indeed very 
close to the experimental value $\gamma_{\rm exp}\simeq 5.4$ 
mJ/mol$\cdot$K$^2$, and weakly increases with doping $x$ as in 
experiment.\cite{Sak10} Thus we 
find that almost no polaronic renormalization of the electron mass 
occurs in Sr$_{1-x/2}$Ce$_{x/2}$MnO$_3$ and Sr$_{1-x}$La$_{x}$MnO$_3$. 

Unlike in the $G$-AF phase, the orbital flavors separate in the $C$-AF 
phase into a wide quasi-1D $z$-subband of width $\sim 4t$ and a narrow 
(empty) $x$-subband, similar to the majority spin bands in the $G$-AF 
phase. The former subband gives a characteristic 1D DOS $N_z(\omega)$, 
with two maxima at both band edges. Due to the AF order in the $ab$ 
plane, the separation of the $e_g$ orbital flavors is almost perfect 
here. Large DOS $N_z(\omega)$ at the edge of the quasi-1D band
[Fig. \ref{fig:dos}(b)] is here sufficient for self-trapping of 
electrons\cite{Feh98} within spin polarons, as is doped 
CaMnO$_3$.\cite{Mes04} In the present case we believe that the polarons 
are quasi-1D, in agreement with the nature of underlying $C$-type 
structure and showing a similar behavior to orbital polarons in a 1D 
model of hole-doped manganites.\cite{Dag04} We thus expect a strongly 
anisotropic optical conductivity in the insulating $C$-AF phase: 
$\sigma_{ab}(\omega)$ with a large gap, and a pseudometallic 
$\sigma_c(\omega)$ with small charge gap.

\section{Magnetic phase transition}
\label{sec:phd}

\begin{figure}[t!]
\includegraphics[width=8cm]{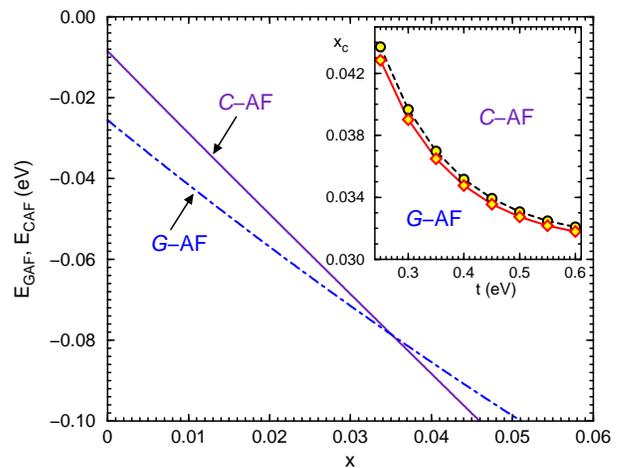}
\caption{(Color online) 
Total energies of the $G$-AF ($E_{G\rm{ AF}}$) and $C$-AF 
($E_{C\rm{ AF}}$) phase as obtained for increasing electron doping $x$. 
Parameters: $t=0.4$ eV, $J_H=0.74$ eV, $g=3$ eV, $K=20$ eV, 
and $J=3.79$ meV. 
Inset shows the phase diagram in the $(t,x_c)$ plane obtained for:
$g=0$ (circles, dashed line), and
$g=3$ eV (diamonds, solid line).
}
\label{fig:phd}
\end{figure}

Now we show that the DE triggers the magnetic transition in an electron 
doped system by comparing the total energies in both magnetic phases. 
In the undoped system ($x=0$) the ground state is the $G$-AF structure, 
with low magnetic energy of $E_{G{\rm AF}}^0=-3JS^2$ per site, and this 
energy increases to $-3JS^2\cos 2\theta$ for finite doping $x>0$. In 
the collinear $C$-AF phase this energy is enhanced to 
$E_{C{\rm AF}}^0=-JS^2$, as the superexchange bonds along the $c$ axis 
are here frustrated. Simultaneously, however, increasing doping gives a 
different kinetic energy in the above phases, obtained by summing up 
the energies of occupied states for a given doping $x$. 
The lower edge of the DOS is much lower in case of the $C$-AF phase 
than for the $G$-AF phase; for instance at $x=0.05$ these minima are at 
$\omega_{\rm G}\simeq -1.63$ eV and 
$\omega_{\rm C}\simeq -2.07$ eV, see Fig. \ref{fig:dos}.
One finds the energy gain due to full electron hopping $\sim t$ along 
the $c$ axis in the 1D electronic structure of the $C$-AF phase, while 
only a small fraction of $t$ contributes to the kinetic energy in the 
isotropic 3D $G$-AF phase. Therefore, the total energy of the $C$-AF 
phase decreases faster with increasing doping $x$, see Fig. 
\ref{fig:phd}, and finally this phase becomes more stable than the 
$G$-AF one for $x>x_c$, as observed in the experiment.\cite{Sak10} 
We note that small spin canting in the $G$-AF phase has almost no 
effect on the phase boundary as the gain in the kinetic energy is 
almost cancelled by the loss of the superexchange energy.

The crucial role played by the DE mechanism is visible in the phase 
diagram of electron doped manganites shown in the inset of Fig. 
\ref{fig:phd}. For increasing hopping $t$ both phases have increased 
kinetic energy gains, but a larger energy gain is found in the $C$-AF 
phase. As a result, the critical concentration $x_c$ is lowered when 
$t$ increases, but one finds that $x_c$ lies still in a relatively 
narrow window of doping $0.03<x_c<0.04$ in a broad range of realistic 
values of $t\in(0.3,0.6)$ eV.

The magnetic transition is first order and occurs at a somewhat lower 
doping $x_c$ when 
the tetragonal distortion which stabilizes the $C$-AF phase is 
considered, see the inset of Fig. \ref{fig:phd}. However, this energy 
gain is rather small compared with the kinetic energy difference 
between the two magnetic phases, so we conclude that the tetragonal 
distortion has only a weak stabilizing effect for the 1D $3z^2-r^2$ 
structure in the $C$-AF phase, while the transition itself follows from 
the DE interaction. We remark that this scenario does not apply to 
La$_x$Ca$_{1-x}$MnO$_3$ nanoparticles where phase separation into AF 
and FM domains takes place.\cite{Wan11}

\section{Summary}
\label{sec:summa}

Summarizing, we have introduced the microscopic model that explains the 
mechanism of the magnetic transition in electron doped manganites from 
canted $G$-AF to collinear $C$-AF phase at low doping $x\simeq 0.04$. 
We demonstrated that the DE supported by the cooperative Jahn-Teller 
effect leads then to dimensional reduction from an isotropic 3D $G$-AF 
phase to a quasi-1D order of partly occupied $3z^2-r^2$ orbitals in the 
$C$-AF phase. This prediction of the theory can be verified by future 
angle resolved photoemission experiments as the shape of the Fermi 
surface in the $G$-AF and $C$-AF phases is radically different. 

The presented theory explains as well the absence of spin canting in the 
$C$-AF phase by the Fermi surface topology. This is a subtle effect as 
electron doping occurs here near the $X$ and $Y$ points in the Brillouin
zone, and the kinetic energy cannot be gained in spin canted structure.

\acknowledgments

We thank Lou-Fe' Feiner and Peter Horsch for insightful discussions.
A.M.O. acknowledges support by the Foundation for Polish Science (FNP)
and by the Polish Ministry of Science under Project N202 069639.



\begin{thebibliography}{99}

\bibitem{Ima98} P. A. Lee, N. Nagaosa, and X.-G. Wen,
                   \rmp \textbf{78}, 17 (2006);
                M. Vojta, Adv. Phys. \textbf{58}, 699 (2009).

\bibitem{Pag10} J. Paglione and R. L. Greene,
                   Nat. Phys. \textbf{6}, 645 (2010).

\bibitem{Dag01} E. Dagotto, T. Hotta, and A. Moreo,
                   Phys. Rep. \textbf{344}, 1 (2001);
                E. Dagotto, New J. Phys. \textbf{7}, 67 (2005).

\bibitem{Tok06} Y. Tokura,
                   Rep. Prog. Phys. \textbf{69}, 797 (2006).

\bibitem{Feh04} A. Wei\ss{}e and H. Fehske,
                   New J. Phys. \textbf{6}, 158 (2004).

\bibitem{Zen51} C. Zener,
                   Phys. Rev. \textbf{82}, 403 (1951).

\bibitem{deG60} P. G. de Gennes,
                   Phys. Rev. \textbf{118}, 141 (1960).

\bibitem{Mae98} R. Maezono, S. Ishihara, and N. Nagaosa,
                   \prb \textbf{58}, 11583 (1998).

\bibitem{Hot00} T. Hotta, A. L. Malvezzi, and E. Dagotto,
                   \prb \textbf{62}, 9432 (2000);
                A. M. Ole\'s and L. F. Feiner,
                   {\it ibid.\/} \textbf{65}, 052414 (2002);
                D. V. Efremov, J. van den Brink, and D. I. Khomskii,
                   Nature Mat. \textbf{3}, 853 (2004).

\bibitem{Don09} S. Dong, R. Yu, J.-M. Liu, and E. Dagotto, 
                   \prl \textbf{103}, 107204 (2009).

\bibitem{Lia11} S. Liang, M. Daghofer, S. Dong, C. \c{S}en, 
                   and E. Dagotto, 
                   \prb \textbf{84}, 024408 (2011).

\bibitem{Jac08} G. Jackeli and G. Khaliullin,
                   \prl \textbf{101}, 216804 (2008).

\bibitem{McQ07} R. J. McQueeney, M. Yethiraj, S. Chang, W. Montfrooij, 
                    T. G. Perring, J. M. Honig, and P. Metcalf, 
                   \prl \textbf{99}, 246401 (2007);
                    \textbf{100}, 069901(E) (2008).

\bibitem{Mot10} Y. Motome and N. Furukawa,
                   \prl \textbf{104}, 106407 (2010);
                S. Kumar and J. van den Brink,
                   {\it ibid.\/} \textbf{105}, 216405 (2010).

\bibitem{Che11} G.-W. Chern and C. D. Batista,
                   \prl \textbf{107}, 186403 (2011).

\bibitem{notecg} In the $C$-AF ($A$-AF) phase the spin order parameter is 
                 staggered in $ab$ planes (along the $c$ axis), while in
                 the $G$-AF phase it is staggered in all three cubic 
                 directions. 

\bibitem{vdB99} J. van den Brink and D. Khomskii,
                   \prl \textbf{82}, 1016 (1999).

\bibitem{Leo10} I. Leonov, D. Korotin, N. Binggeli, V. I. Anisimov, 
                   and D. Vollhardt,
                   \prb \textbf{81}, 075109 (2010).

\bibitem{Fei99} L. F. Feiner and A. M. Ole\'s,
                   \prb \textbf{59}, 3295 (1999).

\bibitem{Zen01} Z. Zeng, M. Greenblatt, and M. Croft,
                   \prb \textbf{63}, 224410 (2001);
                E. N. Caspi, M. Avdeev, S. Short, J. D. Jorgensen,
                   M. V. Lobanov, Z. Zeng, M. Greenblatt, 
                   P. Thiyagarajan, C. E. Botez, and P. W. Stephens, 
                   {\it ibid.\/} \textbf{69}, 104402 (2004).

\bibitem{Tsu10} H. Tsukahara, S. Ishibashi, and K. Terakura,
                   \prb \textbf{81}, 214108 (2010).

\bibitem{Sak10} H. Sakai, S. Ishiwata, D. Okuyama, A. Nakao, H. Nakao, 
                  Y. Murakami, Y. Taguchi, and Y. Tokura,
                  \prb \textbf{82}, 180409 (2010).

\bibitem{Kil99} R. Kilian and G. Khaliullin,
                   \prb \textbf{60}, 13458 (1999).

\bibitem{Dag04} M. Daghofer, A. M. Ole\'s, and W. von der Linden,
                   \prb \textbf{70}, 184430 (2004).

\bibitem{Kov10} A. M. Ole\'s, P. Horsch, G. Khaliullin, and L. F. Feiner,
                   \prb \textbf{72}, 214431 (2005);
                N. N. Kovaleva, A. M. Ole\'s, A. M. Balbashov, 
                   A. Maljuk, D. N. Argyriou, G. Khaliullin, 
                   and B. Keimer,
                   {\it ibid.\/} \textbf{81}, 235130 (2010).

\bibitem{Fle04} G. S. Rushbrooke and P. J. Wood, 
                   Mol. Phys. \textbf{1}, 257 (1958);
                M. Fleck, M. G. Zacher, A. I. Lichtenstein, W. Hanke,
                   and A. M. Ole\'s,
                   Eur. Phys. J. B \textbf{37}, 439 (2004).

\bibitem{Fei05} L. F. Feiner and A. M. Ole\'s,
                   \prb \textbf{71}, 144422 (2005).

\bibitem{Feh98} G. Wellein and H. Fehske,
                   \prb \textbf{58}, 6208 (1998).

\bibitem{Mes04} H. Meskine, T. Saha-Dasgupta, and S. Satpathy,
                   \prl \textbf{92}, 056401 (2004).

\bibitem{Wan11} Y. Wang and H. J. Fan,
                   \prb \textbf{83}, 224409 (2011).

\end{thebibliography}
\end{document}